# On-the-Fly Interrogation of Mobile Passive Sensors from the Fusion of Optical and Radar Data

Ali Hadj Djilani[#$], Dominique Henry[#], Patrick Pons[#$], Herve Aubert[#$]

[#]LAAS-CNRS, France
[$]Toulouse University, France
{ahadjdjila, dhenry, ppons, aubert}@laas.fr

*Abstract* — In this paper a new method based on the fusion of optical and radar data is proposed to detect and remotely interrogate mobile and passive sensors. The sensors are detected in real time by using an optical camera, while their remote reading is carried out on the fly using a Frequency-Modulated Continuous-Wave radar. The proof-of-concept is established from the interrogation of a microfluidic temperature sensor placed on a conveyer belt.

## I. INTRODUCTION

Over the past decades, the combination of radar and camera technologies has been extensively investigated, particularly in the automotive field, where the focus is on obstacle detection and classification [1], multi-object tracking [2], tracking of Micro Aerial Vehicles [3], and automatic traffic surveillance [4]. The integration of these two technologies is based on their distinct and complementary features. The radar provides precise information on the radial distance/range and velocity of detected objects, while the camera offers high spatial resolution, enabling visual details to be accurately distinguished. This combination delivers superior performance to that achievable with each technology used separately [5]. However, to date, the combination of radar and camera technologies has not been used to track and remotely interrogate passive and wireless sensors. Indeed, radar technology has been previously applied alone to detect, identify, and interrogate passive mobile sensors (see, e.g., [6]), but it is not suitable for real-time applications, as it usually requires large memory storage and time-consuming image processing. Furthermore, integrating these technologies with a mobile sensors can significantly reduce the number of sensors needed to monitor an environment, offering a more efficient and cost-effective solution. In this paper, we introduce a data fusion system that combines radar and camera technologies to both detect, localize, and remotely interrogate passive and wireless passive sensors in motion. Indeed, mobile sensors are detected using an optical camera, while their remote reading is carried out on-the-fly from the real-time processing of the raw data delivered by a radar. The proof-of-concept of the propped data fusion technique is established from the optical detection and radar interrogation of a microfluidic temperature sensor moving on a conveyer belt.

The article is organized as follows: Section II reports the architecture of the proposed optical/radar data fusion system. In Section III, first measurement results demonstrating the feasibility of the on-the-fly and in-door tracking of a sensor in motion are reported. Finally, Section IV presents conclusions and prospects in line with the present research work and results.

## II. OPTICAL DETECTION AND RADAR INTERROGATION OF SENSORS

### A. System overview

The proposed remote sensing system for detecting and interrogating passive (battery-less) and wireless sensors consists of three main devices (Figure 1(a)): (1) a Frequency-Modulated Continuous-Wave (FM-CW) radar, (2) an optical camera and (3) a motorized pan-tilt. Both the radar and camera are mounted on the motorized pant-tilt, allowing for elevation and azimuth angular sweeps.

The camera has a pixel count resolution of $N_X \times N_Z$ pixels. The frame rate of the camera (that is, the number of video frames per second) is denoted by $F_{RATE}$. The camera is used here to detect, locate in two-dimensional (2D) space, and identify a sensor in the scene using a specific color label placed on the sensor. Once the sensor is optically located, the main beam of the FM-CW radar is mechanically steered in the optically estimated direction of the sensor by using the pant-tilt. The sensor data are then read by the radar. In the proposed data fusion technique, a moving sensor is thus detected and tracked by the camera and interrogated on the fly by the radar.

The FM-CW radar transmits a chirp in the direction of the sensor with the repetition time $T_C$ and receives the signal backscattered by the sensor. From the transmitted and received signals, a beat frequency spectrum is then determined and the physical quantity of interest (here, a temperature) as well as the radar-to-sensor distance are estimated from echo radar level. The radar is equipped with two antennas: one for transmission ($T_x$), which is characterized by narrow beamwidths in azimuth and elevation (denoted respectively by $\varphi_A$ and $\theta_A$), and another for reception ($R_x$).

The pan-tilt contributes to keep precise tracking of the sensor by adjusting the azimuth angle $\varphi_P$ and elevation angle $\theta_P$ of the radar main beam in the direction of the sensor. The angular speed for azimuthal mechanical steering is denoted by $V_\Phi$, while for elevation, it is denoted by $V_\theta$. In our spherical-coordinate-based 3D coordinate system, the origin (0, 0, 0) is set at the center of the mobile support of the pan-tilt, as shown in Figure 1(a). In this framework, the position $M_n$ of the sensor at time $t_n$ is defined by the coordinates ($\varphi_n$, $\theta_n$, $R_n$), where $\varphi_n$ and $\theta_n$ denote respectively the azimuth and elevation angles of the sensor location, while $R_n$ designates the radar-to-sensor separation distance.

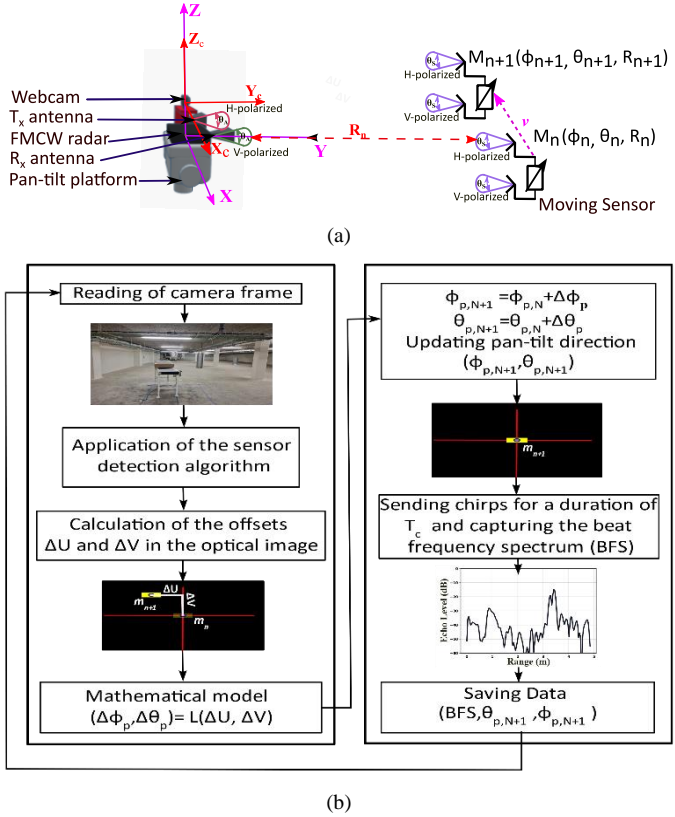

(a)

(b)

Fig. 1. (a) Proposed optical/radar data fusion system for the on-the-fly interrogation of moving sensors; (b) Main steps of the tracking and reading algorithm.

Figure 1(b) illustrates the main steps for acquisition of both optical and radar data. When the sensor moves from position $M_n$ to $M_{n+1}$ with the velocity $v$, the camera captures images of the scene, and from an object detection algorithm based on HSV (Hue Saturation Value) color space [7], the contour of the colored rectangular label attached to the sensor is extracted from the images in the form of a bounding rectangle. The center of coordinates $(U_L, V_L)$ of the rectangle is then derived, providing information on the location of the passive sensor in the image. To estimate the position of the sensor over time, the center $(U_L, V_L)$ of the rectangle is set at the center $C_0$ of the camera image. To do this, we first determine the horizontal and vertical offsets in the optical image, denoted respectively by $\Delta U$ and $\Delta V$, which represent the shifts between the coordinates of the center $(U_L, V_L)$ of the rectangle and the coordinates of center $C_0$ of the image. Next, we apply a linear model to map these spatial offsets into angular offsets $(\Delta \varphi_p, \Delta \theta_p)$ which are ultimately used to control the pan-tilt, to correctly orient the camera and to steer the radar beam in the (optically estimated) direction of the sensor. Then, the radar transmits a chirp in the direction $(\varphi_p, \theta_p)$ with the repetition time $T_C$, receives the signal backscattered by the sensor in that direction and delivers a beat frequency spectrum by mixing the transmitted and received signals. The beat frequency $f_n$ for which the magnitude of the spectral component reaches its highest value is then estimated in real time and used to instantaneously approximate the radar-to-sensor distance $R_n$ by the ratio $f_n c T_c / 4B$, where B and c denote respectively the radar bandwidth and the speed of light in vacuum.

III. EXPERIMENTAL RESULTS AND DISCUSSION

*A. Calibration of the optical/radar data fusion system*

The specific optical/radar data fusion system used to demonstrate the feasibility of the on-the-fly and in-door tracking of a sensor in motion consists of the DK-sR-1030e radar operating at 24 GHz commercialized by *IMST GmbH* [8], a Logitech C270 HD webcam with resolution of 640×480 pixels, and PT-2020 pan-tilt commercialized by *2B Security Systems* (see Figure 2). The transmitting antenna is a horizontally polarized lens antenna with a narrow beamwidth in azimuth and elevation plane, while the receiving antenna is vertically polarized horn (see Table 1 for numbers). The moving sensor is a depolarizing passive and wireless temperature sensor whose radar echo level varies with the temperature. It is a new version of the sensor reported by the authors in [9], as it has been manufactured entirely using three-dimensional metal additive technology. It is composed of a metallic waveguide operating in the frequency bandwidth of the radar and two identical, cross-polarized horn antennas connected through a cavity (see Figure 3). This cavity is crossed at its center by a channel filled with liquid metal (Galinstan): when the microfluidic channel is empty, the transmission coefficient of the cavity is at its maximum; as the volume of liquid metal increases within the channel due to an increase of temperature, the transmission coefficient decreases until reaching its minimum value, corresponding to a microfluidic channel entirely filled with liquid metal. Let $\Delta l$ be the displacement of liquid metal in the channel when the temperature varies by $\Delta T$. The sensitivity $\Delta T / \Delta l$ of the sensor is then given by $S_c / \alpha_v V_0$, where $\alpha_v$ (=11.5×10$^{-5}$ K$^{-1}$) denotes the volumetric thermal expansion coefficient of the liquid metal (Galinstan), $S_c$ (=$\pi r^2$, with r =0.25mm) is the sectional surface of the microfluidic channel, and $V_0$ (=100 mm$^3$) is the tank volume of liquid metal. The sensitivity of the sensor used here is then 17.1 °C/mm. For displacement $\Delta l$ ranging from 0mm to 3.6mm with a step of 0.2mm, the sensor can detect temperature variations up to 61.5 °C with a precision of ±3.4 °C.

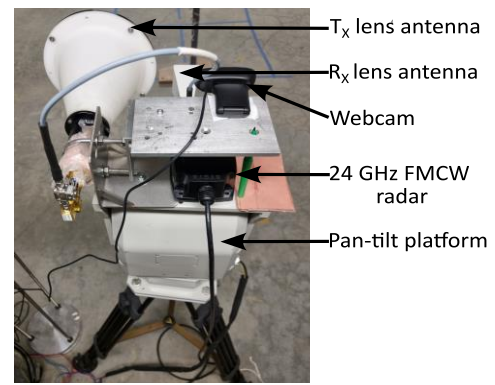

Fig. 2. Photograph optical/radar data fusion system for tracking and reading mobile sensors.

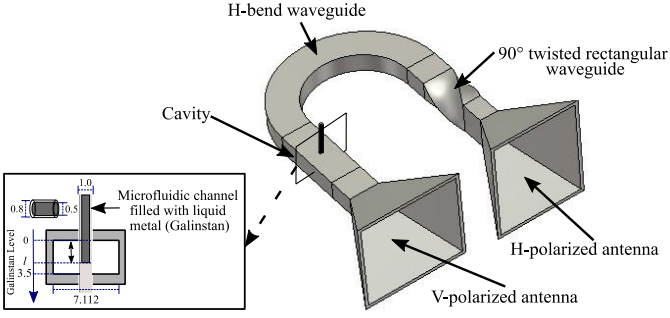

Fig. 3. Picture of the temperature sensor used in the experiment (see [9] for details). Dimensions are in mm.

Table 1. Key descriptors of the optical/radar data fusion and mobile temperature sensors

| Device | Model | Descriptors | |
|---|---|---|---|
| Webcam | Logitech C270 HD | Resolution | 640×480 pixels |
| | | Frame rate $F_{RATE}$ | 1 fps |
| FMCW Radar | DK-sR-1030e from IMST GmbH | Operating frequency | 24 GHz |
| | | Radar bandwidth B | 2 GHz |
| | | $T_x$ antenna gain | 28 dBi |
| | | $R_x$ antenna gain | 20 dBi |
| | | Azimuth beamwidth $\varphi_A$ | 12° |
| | | Elevation beamwidth $\theta_A$ | 9° |
| | | Transmitting Power $P_T$ | 20 dBm |
| | | Number of samples | 256 |
| | | Chirp duration Tc | 15ms |
| | | Range resolution | 7.5 cm |
| Pan-tilt | PT-2020 from 2B Security Systems | Pan Angle $\varphi_P$ | -90° to 90° |
| | | Tilt Angle $\theta_P$ | -60° to 60° |
| | | Precision | 0.1° |
| | | Pan speed $V_\varphi$ | 6.1 °/s |
| | | Tilt speed $V_\theta$ | 4.6 °/s |
| Sensor | Microfluidic temperature sensor [9] | Frequency bandwidth | 22.8-24.8 GHz |
| | | Azimuth beamwidth $\varphi_S$ | 62° |
| | | Elevation beamwidth $\theta_S$ | 50° |
| | | Displacement of liquid metal (Galinstan) in the channel | from 0 to 3.6mm |
| | | Full-scale measurement range of the temperature | 61.5 °C |
| | | Measurement precision of the temperature | ±3.4 °C |

### B. Measurement results and discussion

To demonstrate the feasibility of the on-the-fly and in-door tracking of the temperature sensor in motion, we place the sensor on a 1.5m-long conveyor to obtain linear motion at a constant speed. As illustrated in Figure 4, we perform indoor experiments using two configurations, referred to as *Scenario #1* and *Scenario #2*. In *Scenario #1*, the conveyor is positioned horizontally at 3.5 meters from optical/radar data fusion system (see Figure 4(a)). This configuration allows the sensor to move horizontally at the constant speed of 5 cm/s to track the sensor in the azimuth plane. For *Scenario #2*, the conveyor is positioned diagonally to allow the sensor to move in depth with the speed of 9 cm/s (see Figure 4(b)). This configuration aims to track the sensor as it moves away.

Measurements are performed by varying the level of liquid metal in the microfluidic channel from 0 mm to 2.4 mm, with a step size of 0.2 mm. The temperature is derived from the Galinstan level within the microfluidic channel, with a level of 0 mm indicating a temperature of 20°C.

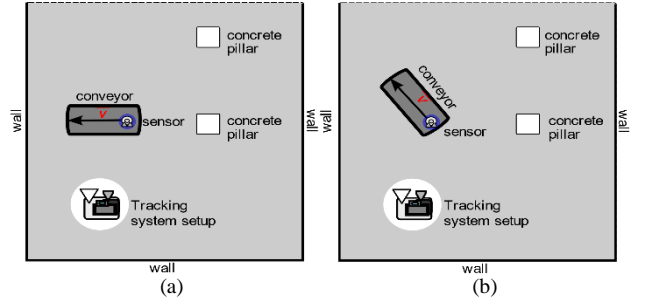

Fig. 4. Experimental indoor configurations for (a) *Scenario #1* and (b) *Scenario #2*.

The measurement results for *Scenario #1* are shown in Figure 5, displaying the radar echo of the sensor as a function of the (optically estimated) azimuth angle φ of the sensor, and for each level of liquid metal in the microfluidic channel. As expected, the radar echo of the sensor decreases as the level of liquid metal increases in the channel. Moreover, the radar echo reaches its highest level for an azimuth angle of 2°, and it decreases as we move away from this angle. This is due to the directional characteristics of the (directive) horn antennas of the sensor. When the Galinstan level exceeds 2.0mm (i.e., for temperature higher than 54.2°), a dispersion of the measured maximum radar echo occurs. This is due to difficulties in isolating the very low radar echo of the sensor from the clutter. Above a Galinstan level of 2.4 mm (i.e., for temperature higher than 61°), this echo is completely embedded in the electromagnetic clutter, making the sensor undetectable. For *Scenario #2*, the results shown in Figure 6 indicate that the radar echo of the sensor decreases with the increasing level of Galinstan in the channel. However, this echo increases as the radar-to-sensor distance increases. This is because initially, the main beam of the receiving antenna of the sensor does not point in the direction of reading system, resulting in a low power received and backscattered by the sensor. As the sensor moves, the main lobes of the two antennas progressively point in the same direction, leading to an increase in the radar echo of the sensor.

Measurement uncertainty of the sensor's radar echo is now determined from using the same interrogation system (radar and antennas) and a motionless sensor. To estimate the measurement uncertainty, we have selected the maximum radar echo of the sensor for different temperature values ranging from 20°C to 61°C, with an azimuth angle set at 3.3° for *Scenario #1* (Figure 7(a)), and a range of 3.8m for *Scenario #2* (Figure 7(b)). These specific values of azimuth angle and range correspond to the highest achievable full-scale measurement range (or dynamic) for both scenarios. In *Scenario #1*, a high full-scale measurement range Δe of 14.4dB on the echo level is obtained with a measurement sensitivity of 0.47dB/°C for temperature ranging from 30.3° to 61°. When the temperature does not exceed 40.5 °C, the standard deviation of the radar echo distribution is less than 0.1dB. Moreover, the standard deviation does not exceed 0.4dB and the average radar echo is less than -18.5dB. For *Scenario#2*, the full-scale measurement range Δe is of 7.4 dB and measurement sensitivity is of 0.25dB/°C. For different temperature variation, the standard deviation of the radar echo distribution does not exceed 0.3dB.

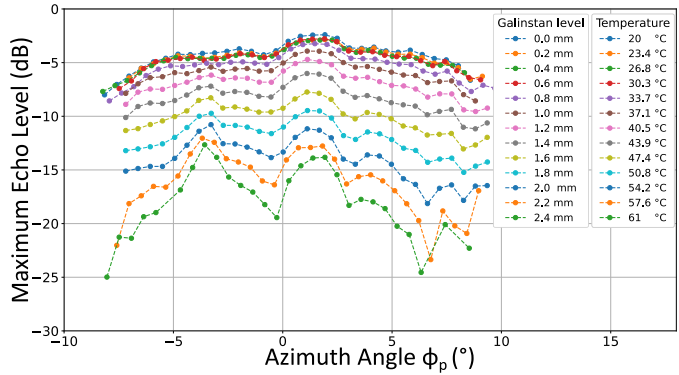

Fig. 5. Echo level of the sensor as a function of the (optically estimated) azimuth of the sensor at various temperatures or equivalently, for different levels of liquid metal (Galinstan) in the channel from 0 mm to 2.4 mm (*Scenario #1*)

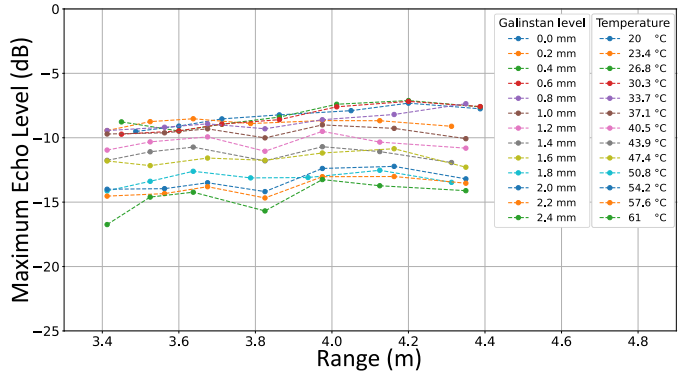

Fig. 6. Echo level of the sensor as a function of the range in the (optically estimated) direction of the sensor at various temperatures or equivalently, for different levels of liquid metal (Galinstan) in the channel from 0 mm to 2.4 mm (*Scenario #2*)

## IV. CONCLUSION

In this paper we proposed optical/radar data fusion system for the on-the-fly and in-door tracking of sensors in motion. The results demonstrate the capability to track and interrogate a temperature sensor effectively. Future work on further improving the proposed system performance includes replacing the current motorized platform with a faster one to track sensor with higher speed (> 9cm/s). In addition, we intend to implement dedicated software to extract 3D trajectory of sensors in real-time, while on-the-fly estimating of the temperature.

ACKNOWLEDGMENT

The authors thank the National Agency for Research (France) for financial support through the S2LAM Project.

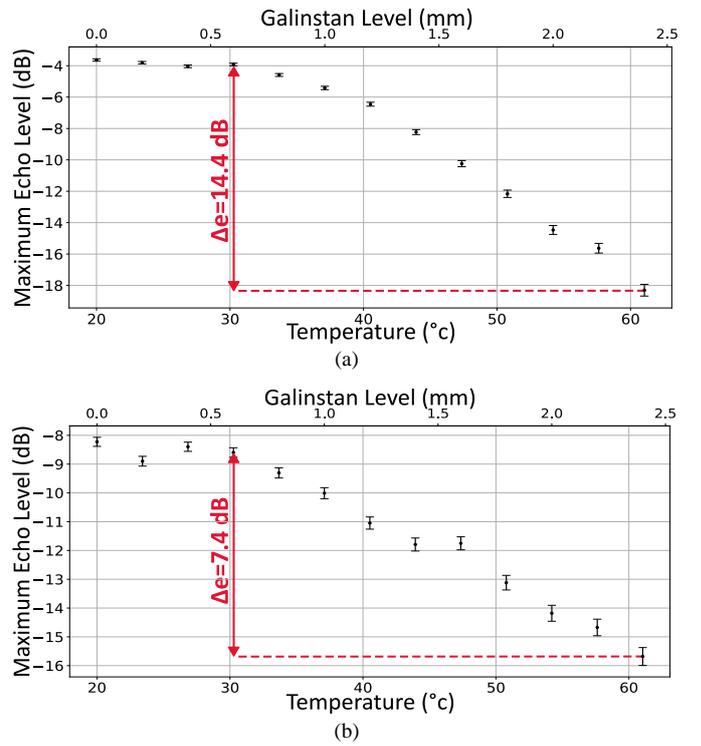

Fig. 7. Echo level of the sensor as a function of temperature (and liquid metal level in the Channel) for (a) Scenario #1 at radar-sensor distance of 3.8m and (b) Scenario #2 at the (optically estimated) azimuth angle of 2.5°. Error bars represent the standard deviation of the radar echo distribution.